\documentclass[a4paper,twoside, 12pt]{article}
\usepackage{setspace}
\usepackage[margin=1in]{geometry}
\usepackage{amsmath,amsfonts,amsthm}
\usepackage{graphicx,cite,times,marvosym} 
\usepackage{enumitem,version,fancyhdr,lipsum}
\usepackage{pdfpages,hyperref}
\usepackage{floatpag}

\newtheorem{example}{Example}

\newtheorem{definition}{Definition}[section]

\numberwithin{equation}{section}
\Large

\begin{document} 
	\title{On the role of the Fibonacci matrix as key in modified ECC}
	\author{Munesh Kumari$^{1}$\footnote{\Letter~ muneshnasir94@gmail.com, \indent 
			{(Orcid: 0000-0002-6541-0284)}
		}, Jagmohan Tanti$^{2}$\footnote{\Letter ~ jagmohan.t@gmail.com}
		\\\normalsize{$^{1}$\small Department of Mathematics, Central University of Jharkhand, India, 835205} 
		\\\normalsize{$^{2}$\small Department of Mathematics, Babasaheb Bhimrao Ambedkar University, Lucknow, India, 226025}    }  
	\date{\today}
	\maketitle
	\noindent\rule{15cm}{.15pt}
	\begin{abstract}
		In this paper, we have proposed a modified cryptographic scheme based on the application of recursive matrices as key in ECC and ElGamal. For encryption, we consider mapping analogous to affine Hill cipher in which a plaintext matrix has been constructed by points corresponding to letters on elliptic curves. In the formation of key-space, the generalized Fibonacci matrices have been taken into account, which is the sequence of matrices. The beauty of considering Fibonacci matrices is their construction where we need only two parameters(integers) in place of $n^2$ elements. The use of a recursive matrix makes a large keyspace for our proposed scheme and increases its efficiency. Thus, it reduces time as well space complexity, and its security \& strength is based on EC-DLP which is a hard problem in number theory. 
		
%
	\end{abstract}
	\noindent\rule{15cm}{.15pt}
	\section{Introduction}
	In discrete mathematics, recursive matrices are rich in many senses like properties, construction, the existence of inverse...etc. One of the popular recursive matrices is generalized Fibonacci matrix \cite{koshy2019fibonacci}, it can be constructed with only two parameters (n,k) and its element is taken from a generalized Fibonacci sequence. Kalika et.al.\cite{prasad2021cryptography} proposed a public key cryptography based on generalized Fibonacci matrices where they also discussed properties of Fibonacci matrices.\\
	Nowadays dependency of the people across the world on the internet has been increased. After arising of COVID-19 situation, most of the meetings, academic activities, business, etc. are taking place in online mode where we used to share a huge number of data (e.g. personal information, video, audio, images, etc.). Now the question arises, is our data received by the authenticated recipient with integrity and confidentiality? So, here comes the Cryptography with the solution to these problems. Cryptography is the science of study about the security, privacy, and confidentiality of information transmitted over a secured channel.\\
	Elliptic curve cryptography(ECC) has recently attracted the industry sectors and academia very much. The reason behind this attraction is that for the equivalent security it uses a significantly smaller key size as compared to other schemes like RSA and ElGmal. Use of ECC reduces processing power, computations, and storage space. Hence it is ideal for implementation.\\
	A public-key cryptosystem based on discrete logarithm problem(DLP) was proposed by T. ElGamal\cite{elgamal1985public} in 1984 which is known as ElGamal Cryptosystem. A new approach to public key cryptography(PKC) using Elliptic curves was taken by N.Koblitz\cite{koblitz1987elliptic} named as Elliptic Curve Cryptography(ECC). The most fascinating thing about ECC was that it uses a smaller key size for the same level of security as the RSA cryptosystem provides. The basics of elliptic curves, algebra on it, and the benefits of using elliptic curve cryptography over RSA can be found in \cite{stinson2005cryptography,hoffstein2008introduction,paar2009understanding,singh2016critical}. Balamurugan et.al \cite{balamurugan2014enhancing} proposed a new approach to the ECC based on the application of matrices and ElGamal technique, further they claimed to provide high security for the encrypted messages. Singh et.al\cite{singh2014new} have proposed an alternative to the problem of a large key size that requires high computing power devices. A fast mapping method based on the matrix approach is presented by Geetha et.al\cite{geetha2014implementation} for ECC, where they offer high security for the encrypted message. Priyatharsan et.al. in \cite{priyatharsan2017new} proposed a new cryptographic system using the elliptic curve cryptography based on square matrices. In\cite{balamurugan2014enhancing,geetha2014implementation,priyatharsan2017new}, they have worked with matrix approach and in this one has to check the invertibility of a matrix and then find the inverse of that matrix which takes much time as well as space. In most of the cryptographic schemes, authors have considered matrix for encryption and key generation. Here comes the requirement of fast computation and key generation, so for that purpose, we are using generalized Fibonacci matrices which help us in fast computation.
      
	This paper is organized as follows. In section 2, the basic concepts of Elliptic Curve Cryptography and properties of multinacci matrices are outlined. In section 3, we have proposed an encryption scheme analogous to affine Hill cipher in which a plaintext matrix has been constructed by points corresponding to letters followed by an example. And lastly, in section 4, we have done analysis on key-space and probability of retrieving the key matrix in the context of $GL_{n}(\mathbb{F}_{p})$.
	\section{Preliminaries}
	\subsection{Elliptic Curve over a finite field \cite{stallings2017cryptography,hankerson2004guide,silverman1992rational}}
	\begin{definition}[Elliptic Curve]
		For $p\in \mathbb{Z}^{+}$ a prime and $\mathbb{F}_{p}= \mathbb{Z}/p\mathbb{Z}$, an elliptic curve $E$ over $\mathbb{F}_{p}$ is the set of points (x, y) with x,y $\in \mathbb{F}_{p}$ which satisfy the equation
		\begin{equation}\label{ecceqn}
			y^{2} = x^{3} +ax +b,
		\end{equation}
		where $a,b \in \mathbb{F}_{p}$ such that $4a^{3} +27b^{2} \ne 0 \pmod p$ together with a single element denoted by $\mathcal{O}$ and called the "point at infinity".
	\end{definition}
	The elliptic curve over $\mathbb{F}_{p}$ is denoted by $E_{\mathbb{F}_{p}}(a,b)$.

	The points on $E_{\mathbb{F}_{p}}(a,b)$ forms an additive abelian group with '$\mathcal{O}$' as the identity element.
	
	\subsection{Generalized Fibonacci Matrix\cite{prasad2021cryptography}}
	\begin{definition}
		The generalized Fibonacci sequence of order $n$ is defined by the recurrence relation, 
		\begin{equation}\label{n^{th}term}
			t_{k+n}=t_{k}+t_{k+1}+t_{k+2}+...+t_{k+n-1}, \hspace{1cm}k\geq 0
		\end{equation}
		with $t_{0}=t_{1}=...= t_{n-2}=0$ and $t_{n-1}=1$.
	\end{definition}
	The generalized Fibonacci sequence is also known as multinacci sequence and it's corresponding matrix is generalized Fibonacci matrix (sometime it is also called multinacci matrix).\\
   Multinacci sequence can be also defined in negative direction by rearranging eqn.(\ref{n^{th}term}), which is given as,
	\begin{equation}
		t_{-k}=t_{-k+n}-(t_{-k+1}+...+ t_{-k+n-1}), \hspace{1cm} for~~ k\geq -1,
	\end{equation}
     where initial values are same as of eqn.(\ref{n^{th}term}).
	\\In particular for $n=2$, eqn.(\ref{n^{th}term}) is Fibonacci sequence [A000045] while for $n=3$, eqn.(\ref{n^{th}term}) gives tribonacci sequence \cite{johnson2008fibonacci,koshy2019fibonacci}.
	\subsubsection*{Multinacci Matrix/Generalized Fibonacci matrix(GFM)}
	The generalized Fibonacci matrix(GFM)\cite{prasad2021cryptography} $F_n^{k}$ of order n is given by,
	\begin{equation}\label{nthpower}	
		F_{n}^{k}=
		\begin{bmatrix}
			t_{k+n-1} & t_{k+n-2}+t_{k+n-3}+...+t_{k} & \cdots &  t_{k+n-2} &\\
			t_{k+n-2} & t_{k+n-3}+t_{k+n-4}+...+t_{k-1} & \cdots & t_{k+n-3} \\ 
			\vdots & \vdots & \ddots & \vdots\\
			t_{k+1} & t_{k}+t_{k-1}+...+t_{k-n+2} & \cdots & t_{k} \\
			t_{k} & t_{k-1}+t_{k-2}+...+t_{k-n+1} & \cdots & t_{k-1}  
		\end{bmatrix}_{n \times n}, ~~~\text{for}~~ k=0, \pm1, \pm2,...
	\end{equation}
	with initial matrix,
	\begin{eqnarray}
		F_{n}=	
		\begin{bmatrix}	
			1 & 1 & 1 & ...& 1 & 1\\	
			1 & 0 & 0 & ...& 0 & 0 \\
			\vdots & \vdots & \vdots & \ddots & \vdots & \vdots  \\
			0 & 0 & 0 & ...& 0 & 0 \\
			0 & 0 & 0 &...& 1 & 0 
		\end{bmatrix}_{n \times n},
	\end{eqnarray}
	where $F_{n}^{0}=I_{n}$ gives identity matrix of order n and $t_{k}$ represents the $ k^{th} $ term of multinacci sequence.\\
	The determinant of $F_{n}^{k}$ is $det(F_{n}^{k})$ = $(-1)^{k(n-1)}$ which is non-zero for all $k$, $n\in \mathbb{N}$.\\
	
 And inverse of GFM is given by,
	\begin{equation}\label{invmulti}
		(F_{n}^{k})^{-1}=F_{n}^{-k}=
		\begin{bmatrix}
			t_{-k+n-1} & t_{-k+n-2}+t_{-k+n-3}+...+t_{-k} & ... & t_{-k+n-2}\\	
			
			t_{-k+n-2} & t_{-k+n-3}+t_{-k+n-4}+...+t_{-k-1} & ... & t_{-k+n-3} \\
			
			
			\vdots & \vdots & \ddots & \vdots \\
			
			t_{-k+1} & t_{-k}+t_{-k-1}+...+t_{-k-n+2} &... & t_{-k} \\
			t_{-k} & t_{-k-1}+t_{-k-2}+...+t_{-k-n+1} &... & t_{-k-1} 
		\end{bmatrix}_{n \times n}.
	\end{equation}
 $i.e.$ to calculate inverse of GFM, one doesn't need to follow usual method, simply replace $k$ by $-k$ in (\ref{nthpower}).\\
 This is the main motivation to consider generalized Fibonacci matrices for encryption-decryption which help us in reducing time complexity and able to perform recursively.
	
	\section{Proposed scheme}
	Multinacci matrices are rich in many sense like multiplication and finding inverse. As we know that for fast and secure communication, we need an efficient algorithm which can generate and verify key quickly. So to make more efficient and fast encryption in ECC, we are taking in account multinacci matrices for key generation, for exchange of parameters ElGamal technique\cite{elgamal1985public} and an encryption scheme analogous to affine hill cipher.\\
	Our proposed method is defined in three part,
	\begin{itemize}
		\item  Construction of a public key
		\item  An algorithm for key generation  
		\item  An encryption scheme and decryption scheme
	\end{itemize}
	which are discussed below step by step.	
	  
	\subsubsection*{Construction of public key:}
	Let $y^{2} \equiv x^{3}+ax+b \pmod p$ such that $4a^{3}+27b^{2} \pmod p \neq 0$ be an Elliptic Curve over the finite field $\mathbb{F}_{p}$ denoted by $E_{\mathbb{F}_{p}(a,b)}$ and $G$ be a cyclic subgroup of $\mathbb{F}_{p}$ with a generator point $E$ . For better understanding, let two entities Alice and Bob want to communicate with each other. So, to construct public Key Alice have to do following steps.\\
	Let us consider $S$=\{$F_{n}^{k}|F_{n}$ is a generalized Fibonacci matrix of order n and $k\in \mathbb{N}$\} over finite field $\mathbb{F}_{p}$.
	\begin{enumerate}
		\item First Alice chooses a primitive element (say $\beta$) of $p$ and a secret number $r$ such that $1< r < p-1$.
		\item Calculate, $E_{1} \equiv \beta^{r} \pmod p.$
		\item Now, $(\beta,E_{1})$ is Alice's public key and $(r)$ is her secret key.
	\end{enumerate}
	\subsubsection*{Key Generation and Encryption:}
	Now, to send message $P$ to Alice, Bob first compute encryption key using Alice's public key $(\beta,E_{1})$ and encrypts his message $P$ as follows:
	\begin{enumerate}
		\item Bob chooses a secret number  $e$ such that $1< e < p-1$.
		\item Next, compute 
		\begin{eqnarray}
			a \equiv \beta^{e} \pmod p ,\nonumber\\
			k \equiv E_{1}^{e} \pmod p.\nonumber
		\end{eqnarray}
		\item Compute key matrix \footnote{$F_{n}^{k}$ is a multinacci matrix of order n which can be directly calculated using multinacci sequence not as usual multiplication.} (say$K$) as $K \equiv F_{n}^{k} \pmod p.$
		\item Convert plain text $P$ in the points of $E_{\mathbb{F}_{p}}(a,b)$, say the points are $P_{1},P_2,P_3,...,P_m$. Now construct a matrix of order $n \times n$ as
		\begin{equation*}
			\begin{bmatrix}	
				P_1 & P_{n+1} & ... & P_{2n-1}\\	
				P_2 & P_{n+2} & ...& P_{}\\
				\vdots & \vdots & \vdots  \\
				P_{n} & P_{2n} & ...& P_{n^2}\\
			\end{bmatrix}_{n \times n}.
		\end{equation*}
		\item \textbf{Encryption:} Now encryption of plain text takes place as follow,
		\\$C \equiv \left[K_{n \times n}  (P_{n \times n}+k  E J_{n \times n})\right]   \pmod p $,  where $J= 
		\begin{bmatrix}\label{Jmatrix}	
			1 &  ... & 1\\	
			\vdots &  \ddots & \vdots  \\
			1 &  ... & 1
		\end{bmatrix}_{n \times n}$ .
		\item Now, Bob will send $(a, C)$ to Alice.
	\end{enumerate}
	\subsubsection*{Decryption:}
	After receiving $(a, C)$ from Bob, Alice perform following operations to recover plaintext:
	\begin{enumerate}
		\item Alice with her secret key($r$), first calculate $k$ as,
		\begin{eqnarray}
			k \equiv a^{r} \pmod p.\nonumber
		\end{eqnarray}
		\item Compute decryption key (say D), 
		\begin{eqnarray}
			D \equiv F_{n}^{-k} \pmod p\nonumber
		\end{eqnarray}
		Where $ F_{n}^{-k} \pmod p$ can be directly calculated (see equation(\ref{invmulti})) with multinacci sequence.
		\item Calculate '$-E$' where $-E$ is a point with the same x-coordinate but negative y-coordinate of $E$ $i.e.$ if $E=(x,y)$, then $-E= -(x,y)=(x,-y)$ which also lies on $E_{\mathbb{F}_{p}}(a,b)$.
		\item Decrypt the ciphertext as
		\begin{eqnarray}
			P &\equiv& (D_{n \times n}  C_{n \times n} +k (-E) J_{n \times n} ) \pmod p\nonumber\\
			&\equiv& (D_{n \times n}  C_{n \times n} -k E J_{n \times n} ) \pmod p\nonumber
		\end{eqnarray}
	where $J$ is given by (\ref{Jmatrix}). 
	\end{enumerate}
	\subsection{Algorithm}
	Let E be a generator point of a cyclic subgroup of $E_{\mathbb{F}_{p}(a,b)}$ and $S$=\{$F_{n}^{k}|F_{n}$ is a generalized Fibonacci matrix of order n and $k\in \mathbb{N}$\} over finite field $\mathbb{F}_{p}$.
	\subsubsection*{Algorithm for Generating Public Key}
	\begin{enumerate}
		\item $\beta$ $ \leftarrow $ a primitive element of $p$ . 
		\item \textbf{Secret Key(r)} $ \leftarrow $ $r$ such that $1< r < p-1$.
		\item \textbf{Public Key:} $ E_{1} \leftarrow \beta^{r} \pmod p.$
		\item Publish $(\beta,E_{1})$ publicly.
	\end{enumerate}
	\textbf{Encryption Algorithm}
	\begin{enumerate}
		\item \textbf{Private Key($e$)} $ \leftarrow e$ such that $1< e < p-1$. 
		\item   $ a \leftarrow  \beta^{e} \pmod p$ and   $ k \leftarrow E_{1}^{e} \pmod p.$
		\item \textbf{Key Matrix:} $K \leftarrow  F_{n}^{k} \pmod p.$
		\item \textbf{Encryption:} $ C \leftarrow  \left[K_{n \times n} (P_{n \times n}+k \times E J_{n \times n})\right] \pmod p $, ,where $J$ is given by (\ref{Jmatrix}).										
		\item \textbf{Transfer} $(a, C)$.
	\end{enumerate}
	\textbf{Decryption Algorithm}\\
	After receiving $(a, C)$,
	\begin{enumerate}
		\item $ k  \leftarrow  a^{r} \pmod p.$
		\item \textbf{Decryption Key:} $ D \leftarrow  F_{n}^{-k} \pmod p$.
		\item \textbf{Decryption:}  $ P  \leftarrow 
		(D_{n \times n} C_{n \times n} -k\times E J_{n \times n} ) \pmod p$.
	\end{enumerate}
	\subsection{Example}
	\begin{example}
		Encrypt the plaintext \textbf{COVID-19} with above proposed scheme.
	\end{example}
	\begin{proof}[Solution:]
		Let $y^{2}=x^{3}+3x+41 \pmod {47}$ be a elliptic curve. We denote it as $E_{\mathbb{F}_{47}}(3,41)$ and $|E_{\mathbb{F}_{47}}(3,41)|=47$.\\
		Now, points of $(E_{\mathbb{F}_{47}}(3,41))$ are :\\
		\begin{tabular}{|c|c|c|c|c|c|c|c|c|c|c|}
			\hline 
			(2,14) & (2,33) & (10,15) & (10,32) & (11,18) & (11,29) & (13,16) & (13,31) & (14,17) & (14,30) \\ 
			\hline 
			(16,7) & (16,40)  & (19,1) & (19,46) & (20,8) & (20,39) & (21,23 & (21,24) & (24,5) & (24,42) \\ 
			\hline
			(27,21) & (27,26) & (28,9) & (28,38) & (30,23) & (30,24) &(33,13) & (33,34) & (34,22) & (34,25) \\
			\hline
			(35,4) & (35,43) & (38,15) & (38, 32) & (40,10) &(40,37) & (41,18) & (41,29) & (42,18) & (42,29) \\
			\hline
			(43,23) & (43,24) & (45,11) & (45,36) & (46,15) & (46,32) & $\mathcal{O}$ & & &\\
			\hline
		\end{tabular}\vspace{.3cm}\\
		Let $E=(2,14)$ be the generator point  of $E_{\mathbb{F}_{47}}(3,41)$.
		We map the letters, symbols, numbers, and special characters as follows:\\
		\resizebox{\linewidth}{!}
		{
			\begin{tabular}{|c|c|c|c|c|c|c|c|c|c|c|c|c|c|}
				\hline
				A  & B  & C  & D  & E & F & G & H & I & J & K & L \\
				(2,14) & (28,9) & (21,24) & (33,34) & (40,10) & (11,29) & (42,29) & (45,11) & (27,26) & (35,4) & (46,15) & (20,39) \\
				\hline
				M & N & o & P & Q & R & S & T & U & V & W & X  \\
				(41,18) & (16,40) & (43,24) & (10,15) & (24,42) & (30,23) & (19,46) & (38,15) & (14,17) & (34,25) & (13,16) & (13,31) \\
				\hline
				Y & Z & 0 & 1 & 2 & 3 & 4 & 5 & 6 & 7 & 8 & 9 \\
				(34,22) & (14,30) & (38,32)  & (19,1) & (30,24) & (24,5) & (10,32) & (43,23) & (16,7) & (41,29) & (20,8) & (46,32)\\
				\hline
				\textasciitilde & ! & @ & \# & \$ & \% & \^{} & \& & * & - & , &  \\
				(35,43) & (27,21) & (45,36) & (42,18) & (11,18) & (40,37) & (33,13) & (21,23) & (28,38) & (2,33) & $\mathcal{O}$ & \\
				\hline
		\end{tabular}}
		\subsubsection*{Construction of Public key:}
	 Consider $S$=\{$F_{2}^{k}| k \in \mathbb{N}$\} over finite field $\mathbb{F}_{47}$.
		\begin{enumerate}
			\item  Alice chooses a primitive element ($\beta$) 31 and her secret key ($r$) 14.
			\item Alice's public key 
			\begin{eqnarray}
				E_{1} \equiv \beta^{r} \pmod p
				\equiv 
				31^{14} \pmod {47}  
				\equiv 
				37 \pmod {47}. 
			\end{eqnarray}
			Here, Alice publish $(\beta,E_{1})$=$(31,37)$ publicly and $r=14$ is her secret key.
		\end{enumerate}
		\subsubsection*{Key Generation and Encryption:}
		Using Alice's public key $(\beta,E_{1})$=$(31,37)$, Bob will generate encryption key and then encrypts his plaintext as follows:
		\begin{enumerate}
			\item Bob chooses his private key such that $1< e < 46$, say $e=21$ .
			\item Next, compute 
			\begin{eqnarray}
				a &\equiv& \beta^{e} \pmod p \equiv 31^{21} \pmod {47} \equiv 38 \pmod {47},\nonumber\\
				k &\equiv& E_{1}^{e} \pmod p \equiv 37^{21} \pmod {47} \equiv 8 \pmod {47}.\nonumber
			\end{eqnarray}
			\item Now, calculate key matrix
			\begin{eqnarray}
				K &\equiv& F_{2}^{8} \pmod {47}
				\equiv 
				\begin{bmatrix}
					f_{9} & f_{8} \\
					f_{8} & f_{7} \\
				\end{bmatrix} \pmod {47}\nonumber\\
				&\equiv&
				\begin{bmatrix}
					34 & 21 \\
					21 & 13 \\
				\end{bmatrix} \pmod {47}.\nonumber
			\end{eqnarray}
			\item 
			Given plaintext is \textbf{'COVID-19'}.\\
			Now mapping letters of plaintext to the point of $(E_{\mathbb{F}_{47}}(3,41))$ which is given by,\\\\
			\begin{tabular}{|c|c|c|c|c|c|c|c|c|c|c|c|c|c|}
				\hline
				C  & O  & V  & I  & D & - & 1 & 9 \\
				\hline
				(21,14) & (43,24) & (34,25) & (27,26) & (33,34) & (2,33) & (19,1) & (46,32) \\
				\hline
			\end{tabular} \vspace{.3cm}
		\item Now constructing plaintext matrices,\\ 
		$P1= 	
			\begin{bmatrix}
				C & V \\
				O & I \\
			\end{bmatrix}$ 
			=$\begin{bmatrix}
				(21,24) & (34,25) \\
				(43,24) & (27,26) \\
			\end{bmatrix}$ and 
			$P2= 	
			\begin{bmatrix}
				D & 1\\
				- & 9\\
			\end{bmatrix}$ 
			=$\begin{bmatrix}
				(33,34) & (19,1)\\
				(2,33) & (46,32)\\
			\end{bmatrix}.$\\
			At last encryption takes place as, 
			\begin{eqnarray}
				C1 &\equiv& \left[ K (P1+k  E J)\right] \pmod p \nonumber\\
				&\equiv&
				\begin{bmatrix}
					34 & 21 \\
					21 & 13 \\
				\end{bmatrix}
				\left( \begin{bmatrix}
					(21,24) & (34,25) \\
					(43,24) & (27,26) \\
				\end{bmatrix} 
				+(45,11)\begin{bmatrix}
					1 & 1 \\
					1 & 1 \\
				\end{bmatrix}\right)\pmod{47}\nonumber\\
				&\equiv&
				\begin{bmatrix}
					(46,15) & (16,40) \\
					(41,18) & (40,10) \\
				\end{bmatrix} \pmod {47} 
				\equiv 
				\begin{bmatrix}
					K & N \\
					M & E \\
				\end{bmatrix}. \nonumber
			\end{eqnarray}
			\begin{eqnarray}
				C2 &\equiv& \left[ K (P2+k  E J)\right] \pmod p \nonumber\\
				&\equiv&
				\begin{bmatrix}
					34 & 21 \\
					21 & 13 \\
				\end{bmatrix}
				\left( \begin{bmatrix}
					(33,34) & (19,1)\\
					(2,33) & (46,32)\\
				\end{bmatrix} 
				+(45,11)\begin{bmatrix}
					1 & 1 \\
					1 & 1 \\
				\end{bmatrix}\right) \pmod {47} \nonumber\\
			&\equiv&
				\begin{bmatrix}
					(27,21) & (16,7)\\
					(16,40) & (20,39)\\
				\end{bmatrix} \pmod {47} \equiv
				\begin{bmatrix}
					! & 6\\
					N & L\\
				\end{bmatrix}. \nonumber
			\end{eqnarray}
			Thus $C=\begin{bmatrix}
				K & N &	! & 6\\
				M & E & N & L\\
			\end{bmatrix}$.\\\\
			Here, Plaintext($P$) = $COVID-19 \rightarrow KMNE!N6L$ = Ciphertext($C$).
			\item Now, Bob can share $(a, C)=(38,15,KMNE!N6L)$ with Alice.\\
		\end{enumerate}
		\subsubsection*{Decryption:}
		After receiving $(a, C)=(38, KMNE!N6L)$ from Bob, Alice perform following operations to recover plaintext:
		\begin{enumerate}
			\item Alice calculates, 
			\begin{eqnarray}
				k &\equiv& a^{r} \pmod p \equiv 38^{14} \pmod {47} \equiv 8 \pmod {47}.\nonumber
			\end{eqnarray}
			\item Calculate Decryption Key$(D)$,
			\begin{eqnarray}
				D &\equiv& F_{2}^{-8} \pmod {47} 
				\equiv 
				\begin{bmatrix}
					f_{-7} & f_{-8} \\
					f_{-8} & f_{-9} \\
				\end{bmatrix} \pmod {47}\nonumber\\
				&\equiv&
				\begin{bmatrix}
					13 & 26 \\
					26 & 34 \\
				\end{bmatrix} \pmod {47}.\nonumber
			\end{eqnarray}
			
			\item Decryption of ciphertext
			\begin{eqnarray}
				P1 &\equiv& (D C1-k  E  J) \pmod {47} \nonumber\\
				&\equiv&
				\begin{bmatrix}
					13 & 26 \\
					26 & 34 \\
				\end{bmatrix}.
				\begin{bmatrix}
					(46,15) & (16,40) \\
					(41,18) & (40,10) \\
				\end{bmatrix} \nonumber
				+ \begin{bmatrix}
					(45,36) & (45,36)\\
					(45,36) & (45,36) \\
				\end{bmatrix} \pmod {47}\nonumber\\
				&\equiv& 
				\begin{bmatrix}
					(21,24) & (34,25) \\
					(43,24) & (27,26) \\
				\end{bmatrix} \pmod {47} \equiv
				\begin{bmatrix}
					C & V \\
					O & I \\
				\end{bmatrix}.\nonumber
			\end{eqnarray} 
			\begin{eqnarray}
				P2 &\equiv& (D C2-k E J) \pmod {47} \nonumber\\
				&\equiv&
				\begin{bmatrix}
					13 & 26 \\
					26 & 34 \\
				\end{bmatrix}.
				\begin{bmatrix}
					(27,21) & (16,7)\\
					(16,40) & (20,39)\\
				\end{bmatrix} \nonumber
				+\begin{bmatrix}
					(45,36) & (45,36) \\
					(45,36) & (45,36) \\
				\end{bmatrix} \pmod {47}\nonumber\\
				&\equiv& 
				\begin{bmatrix}
					(33,34) & (19,1)\\
					(2,33) & (46,32)\\
				\end{bmatrix} \pmod {47} \equiv
				\begin{bmatrix}
					D & 1\\
					- & 9\\
				\end{bmatrix}.\nonumber
			\end{eqnarray} 	
			Thus $P=\begin{bmatrix}
				C & V & D & 1\\
				O & I & - & 9\\
			\end{bmatrix}$.\\\\
			So, KMNE!N6L $\rightarrow$ COVID-19.
		\end{enumerate}
	\end{proof}
	\section{Analysis of Key space }	
	In group theory, general linear group $GL_{n}(\mathbb{F}_{p})$ denotes the group of invertible matrices of order $n$ over the finite field $\mathbb{F}_{p}$ where p is prime and it's order\cite{dummit2004abstract} is given by
	\begin{equation}\label{GLN}
	|GL_{n}(\mathbb{F}_{p})| = (p^{n}-p^{n-1})(p^{n}-p^{n-2}) \cdots (p^{n}-p^{1})(p^{n}-1).
	\end{equation} 
	In our work, we consider the set of generalized Fibonacci matrices over finite field $\mathbb{F}_{p}$ and these matrices are invertible. Here we have presented a table of possible key-spaces for our proposed scheme based on general linear group over a finite field $\mathbb{F}_{p}$ in particular for $n=3$ and $n=4$.
 
		\begin{table}[h]
			\centering
		\begin{tabular}{|c|c|c|c|}
			\hline 
			Prime(p) & Possible Key spaces on $GL_{3}(\mathbb{F}_{p})$ & Possible Key spaces on $GL_{4}(\mathbb{F}_{p})$ \\
			\hline \hline
			29 & 1.3989$\times 10^{13}$ & 2.4131$\times 10^{23}$\\
			\hline
			31 & 2.5559$\times 10^{13}$ & 7.0320$\times 10^{23}$\\ 
			\hline
			37 & 1.2635$\times 10^{14}$ & 1.1995$\times 10^{25}$\\
			\hline
			41 & 3.1920$\times 10^{14}$ & 6.2166$\times 10^{25}$\\ 
			\hline
			43 & 4.9063$\times 10^{14}$ & 1.3336$\times 10^{26}$\\
			\hline
			47 & 1.0948$\times 10^{15}$ & 5.5465$\times 10^{26}$\\
			\hline
			53 & 3.2363$\times 10^{15}$ & 3.8017$\times 10^{27}$\\ 
			\hline
			59 & 8.5136$\times 10^{15}$ & 2.1187$\times 10^{28}$\\
			\hline
			61 & 1.1499$\times 10^{16}$ & 3.6139$\times 10^{28}$\\
			\hline
			67 & 2.6794$\times 10^{16}$ & 1.6239$\times 10^{29}$\\
			\hline		
			\vdots & \vdots & \vdots \\ 
			\hline
		\end{tabular}
		\caption{Possible key spaces of order 3 \& 4.}
	\end{table}
	From table we conclude that increase in value of $p$ results a large key space. Simultaneously, increase in size of matrix also corresponds to a large key space which can be easily observed in above table. Thus, it is impractical to recover the key by an intruder by brute force attack. In such case, intruder may try some other attacks.
	\subsection*{Brute Force Analysis}
  An intruder can not uncover plaintext until and except if, he has not private key for the beneficiary. Also in case of large prime number, the keyspace will be quite large which makes a brute force for finding private key impossible for a intruder. In the following table, we have shown the probability of retrieving encryption key matrix in particular for $n=3~\&~n=4$ by brute force.
	
	\begin{table}[h]
		\centering
		\begin{tabular}{|c|c|c|c|}
			\hline 
			Prime(p) & Probability on $GL_{3}(\mathbb{F}_{p})$ & Probability on $GL_{4}(\mathbb{F}_{p})$ \\
			\hline \hline
			29 & $7.15 \times 10^{-14}$ & $4.14\times 10^{-24}$\\
			\hline
			31 & $3.91 \times 10^{-14}$ & 1.42$\times 10^{-24}$\\ 
			\hline
			37 & 7.91$\times 10^{-15}$ & 8.33$\times 10^{-26}$\\
			\hline
			41 & 3.13$\times 10^{-15}$ & 1.60$\times 10^{-26}$\\ 
			\hline
			43 & 2.03$\times 10^{-15}$ & 7.50$\times 10^{-27}$\\
			\hline
			47 & 9.13$\times 10^{-16}$ & 1.80$\times 10^{-27}$\\
			\hline
			53 & 3.09$\times 10^{-16}$ & 2.63$\times 10^{-28}$\\ 
			\hline
			59 & 1.17$\times 10^{-16}$ & 4.72$\times 10^{-29}$\\
			\hline
			61 & 8.70$\times 10^{-17}$ & 2.77$\times 10^{-29}$\\
			\hline
			67 & 3.73$\times 10^{-17}$ & 6.16$\times 10^{-30}$\\
			\hline		
			\vdots & \vdots & \vdots \\ 
			\hline
		\end{tabular}
		\caption{Probability of retrieving key matrix.}
	\end{table}
 	From the table, we observe that if we take parameters $n$ and $p$ sufficiently large, the probability of getting the key matrix is negligible which makes this cryptographic scheme secure against brute force attack.
	
	
	Some times intruder performs chosen-plaintext attack, where he used to choose a plain text and try to get corresponding ciphertext. With the help of that information intruder try to get encryption key and other relevant information regarding encryption. \\
	In our proposed method, ciphertext C is given by
	\begin{equation}\label{DLP}
		C \equiv K(P+kEJ) \pmod p. 
	\end{equation}
	In equation (\ref{DLP}) it is almost impossible to calculate K even though C and P is known because of the elliptic curve discrete logarithm problem (ECDLP)(see \cite{hankerson2004guide}). So in this context again our proposed methodology is secure against CPA.
	
	\section{Conclusion}
	In this paper, we have proposed a modified public key cryptography along with Elliptic curve cryptography and Elgamal using multinacci matrices over a finite field $\mathbb{F}_{p}$. With the help of multinacci matrices, we reduce time as well as space complexity of computation. We have used recursive nature of direct calculation of inverse matrix and $n^{th}$ power of a matrix for fast computation. Also, we have observed that if we increase the prime number, keyspace increases as well and size of key matrix do not follow primes which results randomness in size. The security of our proposed cryptography depends upon the discrete logarithm problem of ECC which is a hard problem in number theory. In our analysis, we found that our proposed method has large key space, quite robust, and can be implemented easily.
	\subsection*{Acknowledgment}      
	The authors are thankful to anonymous reviewer for their care and advice. The first author acknowledge the University Grant Commission(UGC), India for providing fellowship for this research work.
	
	\bibliography{References}

\begin{thebibliography}{10}

\bibitem{balamurugan2014enhancing}
{\sc Balamurugan, R., Kamalakannan, V., Rahul, G.~D., and Tamilselvan, S.}
\newblock Enhancing security in text messages using matrix based mapping and
  elgamal method in elliptic curve cryptography.
\newblock In {\em 2014 International Conference on Contemporary Computing and
  Informatics (IC3I)\/} (2014), IEEE, pp.~103--106.

\bibitem{dummit2004abstract}
{\sc Dummit, D.~S., and Foote, R.~M.}
\newblock {\em Abstract algebra}, vol.~3.
\newblock Wiley Hoboken, 2004.

\bibitem{elgamal1985public}
{\sc ElGamal, T.}
\newblock A public key cryptosystem and a signature scheme based on discrete
  logarithms.
\newblock {\em IEEE transactions on information theory 31}, 4 (1985), 469--472.

\bibitem{geetha2014implementation}
{\sc Geetha, G., and Jain, P.}
\newblock Implementation of matrix based mapping method using elliptic curve
  cryptography.
\newblock {\em International Journal of Computer Applications Technology and
  Research 3}, 5 (2014), 312--317.

\bibitem{hankerson2004guide}
{\sc Hankerson, D., Menezes, A.~J., and Vanstone, S.}
\newblock {\em Guide to Elliptic Curve Cryptography}.
\newblock Springer Science \& Business Media, 2004.

\bibitem{hoffstein2008introduction}
{\sc Hoffstein, J., Pipher, J., Silverman, J.~H., and Silverman, J.~H.}
\newblock {\em An introduction to mathematical cryptography}, vol.~1.
\newblock Springer, 2008.

\bibitem{johnson2008fibonacci}
{\sc Johnson, R.~C.}
\newblock Fibonacci numbers and matrices.
\newblock {\em manuscript available at http://www. dur. ac. uk/bob.
  johnson/fibonacci\/} (2008).

\bibitem{koblitz1987elliptic}
{\sc Koblitz, N.}
\newblock Elliptic curve cryptosystems.
\newblock {\em Mathematics of computation 48}, 177 (1987), 203--209.

\bibitem{koshy2019fibonacci}
{\sc Koshy, T.}
\newblock {\em Fibonacci and Lucas numbers with applications}.
\newblock John Wiley \& Sons, 2019.

\bibitem{paar2009understanding}
{\sc Paar, C., and Pelzl, J.}
\newblock {\em Understanding cryptography: a textbook for students and
  practitioners}.
\newblock Springer Science \& Business Media, 2009.

\bibitem{prasad2021cryptography}
{\sc Prasad, K., and Mahato, H.}
\newblock Cryptography using generalized fibonacci matrices with affine-hill
  cipher.
\newblock {\em Journal of Discrete Mathematical Sciences and Cryptography\/}
  (2021), 1--12.

\bibitem{priyatharsan2017new}
{\sc Priyatharsan, U., Rupasinghe, P.~L., and Murray, I.}
\newblock A new elliptic curve cryptographic system over the finite fields.
\newblock In {\em 2017 6th National Conference on Technology and Management
  (NCTM)\/} (2017), IEEE, pp.~164--169.

\bibitem{silverman1992rational}
{\sc Silverman, J.~H., and Tate, J.~T.}
\newblock {\em Rational points on elliptic curves}, vol.~9.
\newblock Springer, 1992.

\bibitem{singh2014new}
{\sc Singh, L.~D., and Debbarma, T.}
\newblock A new approach to elliptic curve cryptography.
\newblock In {\em 2014 IEEE International Conference on Advanced
  Communications, Control and Computing Technologies\/} (2014), IEEE,
  pp.~78--82.

\bibitem{singh2016critical}
{\sc Singh, S.~R., Khan, A.~K., and Singh, T.~S.}
\newblock A critical review on elliptic curve cryptography.
\newblock In {\em 2016 International Conference on Automatic Control and
  Dynamic Optimization Techniques (ICACDOT)\/} (2016), IEEE, pp.~13--18.

\bibitem{stallings2017cryptography}
{\sc Stallings, W.}
\newblock {\em Cryptography and network security: principles and practice}.
\newblock Pearson Upper Saddle River, 2017.

\bibitem{stinson2005cryptography}
{\sc Stinson, D.~R.}
\newblock {\em Cryptography: theory and practice}.
\newblock Chapman and Hall/CRC, 2005.

\end{thebibliography}
	\bibliographystyle{acm}
	
\end{document}